\documentclass[a4paper,fleqn]{cas-sc}
\usepackage{blindtext}
\usepackage{hyperref}
\usepackage{caption}
\usepackage{subcaption}
\usepackage{algorithm}
\usepackage{algpseudocode}

 \usepackage[numbers]{natbib}

\begin{document}
\let\WriteBookmarks\relax
\def\floatpagepagefraction{1}
\def\textpagefraction{.001}

\shorttitle{GPU Filters 2D Convex Hulls}

\shortauthors{Carrasco et~al.}

\title [mode = title]{An Evaluation of GPU Filters for Accelerating the 2D Convex Hull}                      



%

\author[1]{Roberto Carrasco}[]
\address[1]{Departamento Ciencias de la Computaci\'on, Universidad de Chile.}
\address[2]{Instituto de Inform\'atica Universidad Austral de Chile}







\author[2]{H\'ector Ferrada}[]

\author[2]{Crist\'obal A. Navarro}[]


\author[1]{Nancy Hitschfeld}




\begin{abstract}
The Convex Hull algorithm is one of the most important algorithms in computational geometry, with many applications such as in computer graphics, robotics, and data mining. Despite the advances in the new algorithms in this area, it is often needed to improve the performance to solve more significant problems quickly or in real-time processing. This work presents an experimental evaluation of GPU filters to reduce the cost of computing the 2D convex hull. The technique first performs a preprocessing of the input set, filtering all points within an eight-vertex polygon in logarithmic time, to obtain a reduced set of candidate points. We use parallel computation and the use of the Manhattan distance as a metric to find the vertices of the polygon and perform the point filtering. For the filtering stage we study different approaches; from custom CUDA kernels to libraries such as Thrust and CUB. Three types of point distributions are tested: a normal distribution (favorable case), circumference (the worst case), and a case where points are shifted randomly from the circumference (intermediate case). Experimental evaluation shows that the GPU filtering algorithm can be up to $23 \times$ faster than a sequential CPU implementation, and the whole convex hull computation can be up to $30 \times$ faster than the fastest implementation provided by the CGAL library. 

\end{abstract}


\begin{highlights}
    \item An experimental study of filtering approaches for accelerating the 2D convex hull.
    \item Four filtering variants are evaluated; one CUDA kernel, two Thrust functions and one CUB primitive. 
    \item All approaches are compared to the CGAL library which is a standard reference.
    \item Three point distributions are tested; normal, circumference and displaced circumference. 
    \item The source code and benchmarks are available for the community.
\end{highlights}

\begin{keywords}
GPU Computing, Computational Geometry, Convex Hull, Filtering Techniques, Parallel Reduction
\end{keywords}

\maketitle

\section{Introduction}
The convex hull is a geometric concept represented by the smallest convex polygon that encloses a given set of points in the plane \cite{Berg:2008:CGA:1370949}. Informally, it is the shape formed by connecting all the points on the outermost boundary of a set of points. 
The study of convex hulls has a rich history, with many different algorithms and techniques being developed over the years to improve the efficiency and accuracy of convex hull computations \cite{o1998computational}. 

Convex hulls have a wide range of applications in various fields of science and technology. Some examples include:
\begin{itemize}
\item Computer graphics:  convex hulls are used to compute the convex bounding polygons of shapes for accelerating geometric operations over them and for constructing other structures like Delaunay triangulations and Voronoi diagrams\cite{Berg:2008:CGA:1370949,polylla01}, among others.

\item Collisions: Robotics use convex hulls to compute a robot's reachable space, which is the set of points that the robot can move to without colliding with obstacles \cite{MEERAN1997737, Nearchou1994}.

\item Data mining: It uses convex hulls to cluster a set of points into groups based on their spatial relationship. That is useful for discovering patterns or trends in large datasets \cite{NEMIRKO2021381}.

\end{itemize}

Overall, convex hull algorithms have a wide range of applications in many different fields, making them valuable tools for solving a variety of problems.
The convex hull algorithms have advanced significantly over the years, and algorithms have been developed, such as the Gift wrapping  \cite{10.1145/321556.321564,JARVIS197318} in $O(nh)$ time (where $n$ is the number of points in the original set of points, and $h$ is the number of points in the hull), the Graham scan \cite{GRAHAM1972132} in $O(n \log n)$, the QuickHull  \cite{10.1145/235815.235821} in $O(n^2)$ in worse case and $O(n \log n)$ in average, the divide-and-conquer algorithm \cite{10.1145/359423.359430}  and the incremental approach \cite{KALLAY1984197} in $O(n \log n)$, among others. In addition, new algorithms have been developed that can handle special cases, such as computing the convex hull of points on the surface of a sphere or in higher-dimensional spaces. Overall, the state-of-the-art in convex hull computation continues to evolve and improve, offering increasingly efficient and accurate solutions to a wide range of problems. Two of the most efficient implementations of the convex hull are provided by Qhull \cite{10.1145/235815.235821}, and CGAL \cite{cgal:hs-ch3-18b} libraries, which implement one or more of the algorithms mentioned above. It should be noted that CGAL is considered the standard reference to compare with future solutions in this work. Both QHull and CGAL libraries have been optimized for a sequential computation scheme. In the latest years, many applications in computational geometry have moved from sequential single-core CPU computation to parallel computation  \cite{10.1145/3402819} in order to tackle larger problems without incurring long execution times. One of the most common transitions has been moving from CPU to GPU-based computation.

The goal of this work is to study the various programming methods available for modern GPUs that incorporate new cores, and data management libraries that optimize efficiency. The proposed algorithm is implemented in four different variations and subjected to a thorough performance evaluation to determine the impact of these cutting-edge techniques and technologies. 
Ferrada et al \cite{FERRADA2020112298}, recently developed a sequential optimization of the convex hull algorithm, in this research, we develop a way of accelerating the preprocessing stage of the convex hull algorithm, discarding all points inside of an eight-sided polygon, taking advantage of parallelization on the modern GPUs. We also show the experimental results of four variants implemented using the GPU programming model. All the implementations of this work have a special focus on the following key ideas:

\begin{itemize}
    \item Fast preprocessing: the main result of this work is the evaluation of four filtering variants for computing the convex hull of a set of points.
    \item Efficient operations: based on the work of  \cite{FERRADA2020112298}, another key to our contribution is the use of the Manhattan distance as a metric to determine the initial polygon and parallel techniques for computing the preprocessing.
    \item Parallel algorithms: Another key aspect of our contribution is to use the latest GPU programming model \cite{ScanTC,9147055,nvidia2020A100} in implementations.
    \item Scalability: this work presents a kernel implementation with improved scalability, capable of handling larger datasets than those supported by current state-of-the-art libraries. 
\end{itemize}

The rest of the manuscript is organized as follows;  related work is covered in  Section  \ref{sec:related_works}. The problem statement and main contribution are presented in Section \ref{sec:contribution}, and different implementation variants of the algorithm are described in Section \ref{sec:implementation}. The  distributions of points under study are described in Section \ref{sec_study}. An in-depth experimental performance comparison against a faster implementation available in CGAL is presented in Section \ref{sec_experiment}. Finally, a discussion of the results as well as conclusions are given in Section \ref{sec_conclu}.

\section{Related Works}
\label{sec:related_works}
Previous attempts to speed up the computation of convex hulls have employed parallel algorithms to implement some operations of the traditional algorithms, such as calculating the distance between points or determining extreme points, among others, or using parallel preprocessing techniques to select candidate points. Srungarapu et al. proposed a parallel GPU-based QuickHull algorithm to accelerate the computation of 2D convex hulls, they offer a QuickHull-based algorithm that parallelized the determination of the extreme points, marking the points inside of the polygon and scanning but the main loop of the QuickHull is in CPU. They reported a speedup of up to $14 \times$ over a traditional CPU-based convex hull solution \cite{5763404}.

Mei proposed a GPU-based solution \cite{mei} that utilizes a preprocessing approach to classify all points and discard those that do not belong to the convex hull in GPU. This preprocessing step resulted in a speedup of up to $6 \times$ over a Qhull implementation. Additionally, Stein et al. presented a parallel algorithm for computing the 3D convex hull of a set of points using the CUDA programming model \cite{STEIN2012265}. This approach, based on the QuickHull method, achieved $30 \times$ of speedup over a CPU-based Qhull implementation.

Blelloch et al. present a theoretical analysis about a parallel incremental randomized algorithm \cite{10.1145/3350755.3400255} for computing the 2D convex hull that, for a set of $n$ points in any constant dimension, has an $O(\log n)$ dependence depth with high probability. This leads to a simple, work-optimal parallel algorithm with a polylogarithmic span.

Barbay and Ochoa proposed an adaptive algorithm for merging $k$ convex hulls on the plane. The algorithm begins by decomposing the input sequence of points into several parts and calculating the convex hull for each part, both steps can be done in linear time. They then use a novel and fast merge technique to join all the partial hulls. Both of these works demonstrate the potential of using GPU to accelerate convex hull algorithms.

The most widely-used and efficient method for improving computational performance is eliminating interior points not candidates to the hull. However, many algorithms strongly depend on the input size, and preprocessing algorithms can significantly reduce the input size by $O(n)$ discarding points irrelevant to the convex hull algorithm \cite{cudach2d,cudach3d}. Mei has developed a filtering technique for computing the convex hull in 2D and 3D by first identifying $16$ points on the convex hull through rotation of all points in GPU at three different angles and then discarding all points in parallel inside the polygon formed by these identified points. Additionally, Mei et al. propose iterative filtering techniques in GPU to improve the computation of the convex hull further \cite{mei2012}.

One of the most recent works in preprocessing technique for computing the convex hull is the proposal by Alshamrani et al. \cite{alshamrani}, who proposed a filtering technique that uses the Euclidean distance to find the extreme points by filtering all the points within a polygon of four vertices, achieving an acceleration of up to $77 \times$ and $12 \times$ faster than the Graham scan and Jarvis march algorithms, respectively. The other important and recent work is the proposal by Ferrada et al. \cite{FERRADA2020112298}, who developed a sequential approach, named \texttt{heaphull}, for discarding points in 2D convex hulls using the Manhattan distance as the primary metric. This method discards all points outside a polygon formed by eight vertices in $O(n)$, resulting in a reduced set of candidate points. They reported a speedup of $1.7 \times$ to $10 \times$ faster than convex hull methods available in the CGAL library. In this last work, they followed working in the GPU implementation where Alan et al. \cite{alanhull} report $4 \times$ faster than the sequential CPU-based algorithm (\texttt{heaphull}) and $3 \thicksim 4 \times$ faster than other existing GPUs-based approaches in state-of-the-art.

Many of these recent CPU-based works have the potential to be implemented using the GPU programming model, such as the work by Ferrada et al. \cite{FERRADA2020112298}, which could obtain significant improvements in performance by leveraging the parallel processing capabilities of GPUs. This presents an opportunity to enhance current filtering methods by incorporating the GPU programming model and low-cost geometric operations.

\section{Parallel Algorithm for the Convex Hull Filtering}
\label{sec:contribution}

In this section we present a GPU filter technique that builds on the CPU-based work developed by Ferrada et al. \cite{FERRADA2020112298} that utilizes the Manhattan distance as the primary metric for calculating extreme points. The proposed algorithm consists of three stages. The first stage, outlined in (\ref{subsec:extreme_points}), involves constructing an octagon polygon with extreme points from the input. The second stage, described in (\ref{subsec:filter_phase}), discards all points not candidates for the hull. The final stage, outlined in (\ref{subsec:ch_compute}), calculates the convex hull using an existing state-of-the-art algorithm. Finally, we show the complexity of the algorithm.

\subsection{Finding the Eight-side Polygon}
\label{subsec:extreme_points}
\begin{figure}[ht!]
    \centering
    \includegraphics[scale=0.5]{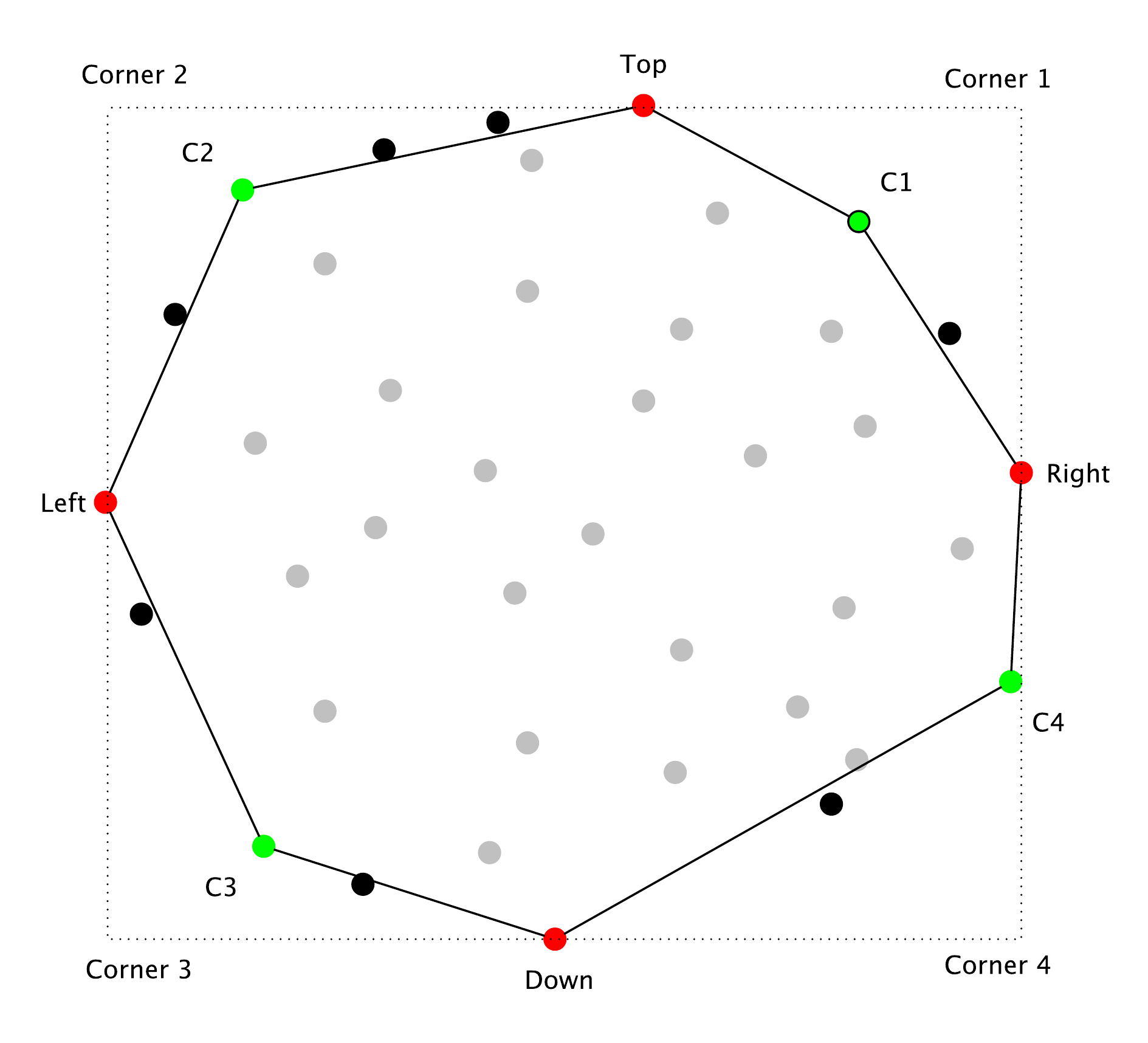}
    \caption{The illustration depicts the eight-sided polygon resulting from the first stage of the algorithm applied to a point cloud. Points colored in black are considered candidates for the hull, while those colored in gray are discarded in the filtering stage. The extreme points for each coordinate axis are represented in red, and the corner points are represented in green.}
    \label{fig:filter}
\end{figure}

The first stage of the algorithm involves constructing an eight-sided polygon from the $n$ input points, which is then used to filter the points and retain the remaining $n'\leq n$ points as convex hull candidates. Figure \ref{fig:filter} illustrates the point classification of the hull candidates resulting from applying this stage, showing the eight-sided polygon, extreme points, and corner points. The polygon is formed by the four extreme points (the right-most, the upper-most, the left-most, and the lowest-most) and the four points that, according to the Manhattan distance, are closest to the corners of the bounding box (see Figure \ref{fig:filter}) defined by the four extreme points. The extreme points are obtained using parallel min-max reduction in each axis in logarithmic time, using a parallel reduction based on the block-warp hierarchy of modern GPU architecture. Meanwhile, each one of the four corner points is the lowest Manhattan distance between each point and the corner formed by the projection of each consecutive pair of axis extreme points (e.g., the top-right corner is the rightmost $x$-coordinate and the topmost $y$-coordinate) in a counterclockwise fashion. This is done in logarithmic time using parallel reduction to compute the minimum distance. All points inside the polygon formed by these points are guaranteed not to belong to the convex hull. The algorithm takes $O(\log n)$ time to find all these points. 

\begin{figure}[htb]
    \centering
    \includegraphics[scale=0.8]{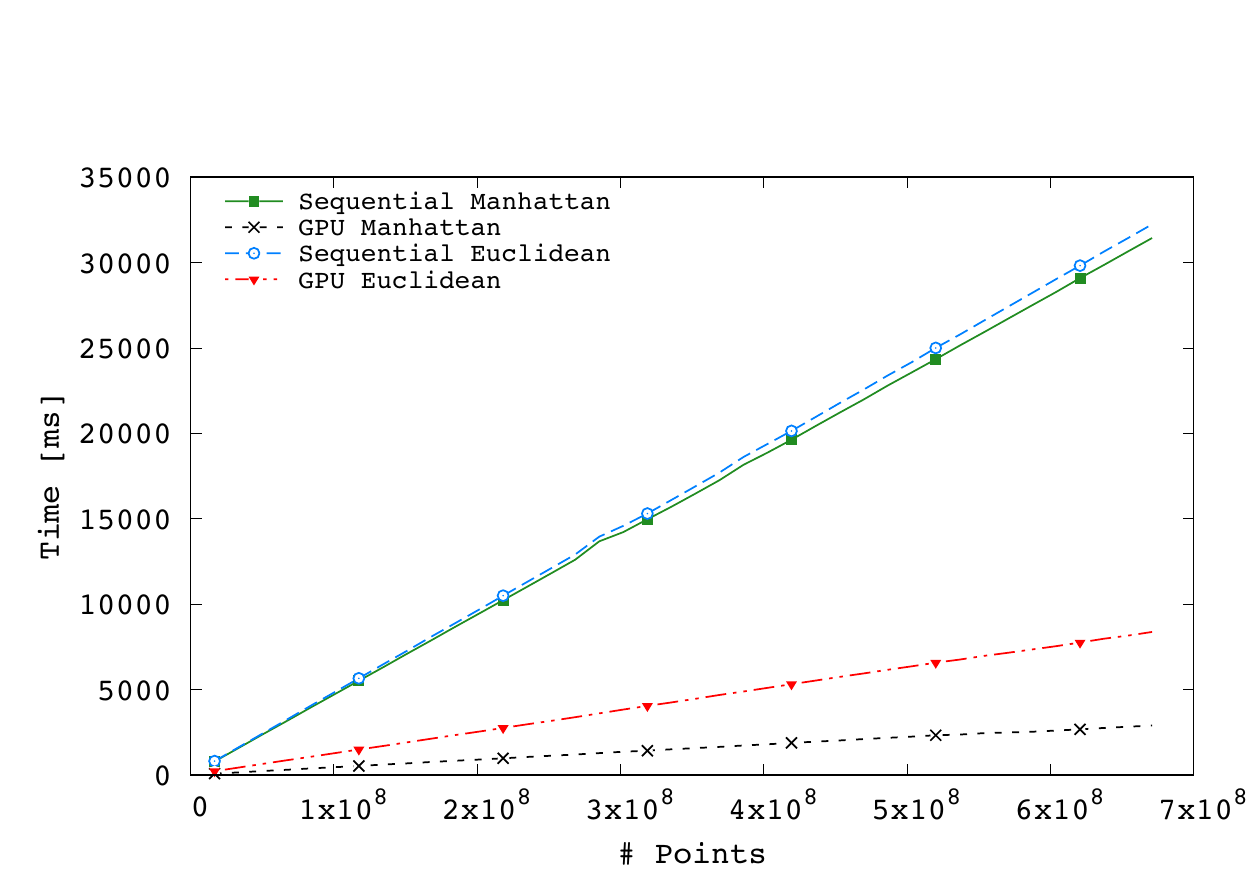}
    \caption{Performance of finding extreme points between Manhattan and Euclidean.}
    \label{plot_mve_circ}
\end{figure}

The aim of this stage is to find the corner points in logarithmic time. 
For the distance function, the Manhattan distance \cite{krause1986taxicab} is used, which finds the point that minimizes the sum of the vertical and horizontal distances to each corner. Additionally, the Manhattan distance is a simple and inexpensive computation that only requires addition and subtraction, whereas the Euclidean distance requires products and computationally expensive operations such as square roots \cite{Deza2009}. Figure \ref{plot_mve_circ} shows the time spent to find all axis extreme points of a circumference (including the axis and corners). In the figure it is possible to observe that the computation time spent in computing the Manhattan distance is less expensive  than the Euclidean distance in both cases, GPU and CPU. 

\subsection{Building the Filter}
\label{subsec:filter_phase}
The next stage of the algorithm involves a GPU kernel call, where each point is processed by a separate thread, and it checks if it lies within the polygon built in the previous stage. If the point is inside the polygon, it is filtered by the algorithm, as can be seen in Figure \ref{fig:filter} where the black points are considered as candidates for the hull, and the gray points are discarded. As a result of this stage, we have a vector of n bits, where each bit set to $1$ represents a candidate point for the hull, and if it is set to $0$, the point does not belong to the hull.

However, it is not possible to compact all the elements of a sparse array simultaneously in a single parallel kernel call, since the position of each element of the original array in the compacted array is not known in advance. Nevertheless, there are techniques to compact sparse arrays in logarithmic time \cite{BlellochTR90}. The most common way to compact a sparse bit array to another non-sparse array of a smaller size is to use the parallel scan or atomic functions, this work uses a custom compaction developed for us, based on the tensor cores scan developed by Dakkak et al. \cite{ScanTC} adapted to work with integer arrays.

\subsection{Computing the Convex Hull}
\label{subsec:ch_compute}
As a result of the previous stage, we have a set of $n'$ candidate points from the initial input point set (red, green, and black points in Figure \ref{fig:filter}, all gray points were discarded in the previous stage). The algorithm can be connected to any existing convex hull implementation available in the state-of-the-art. In this work, we use an efficient CPU implementation provided by CGAL for all experiments, which is described in more detail in Section \ref{sec_experiment}.

\subsection{Complexity of the Algorithm}
The complexity of the algorithm is determined by the sum of the complexities of the first two phases, and the operations carried out in them. Starting with the construction of the eight-sided polygon, the cost of this phase is given by the search for the extreme points, which has a linear cost in sequential and logarithmic cost in parallel. The second phase, which corresponds to the construction of the filter, is logarithmic. This is due to the fact that, unlike its sequential version that has a linear cost, whereas it goes through the points to determine if it is inside the polygon, it immediately writes in an array of least extension. In parallel, this phase consists of two subphases. The first subphase determines if the points are candidates, which only requires one call to the kernel in GPU in $O(1)$, whereas the second subphase, corresponding to the compaction of the array in step, has a logarithmic cost. Consequently, the sum of both phases gives an $O(\log n)$ cost, while their CPU counterparts offer only linear-time solutions.

\section{Filter Implementations}
\label{sec:implementation}
There are various approaches to programming on a GPU that involve different programming incorporating new types of cores and libraries for efficient data management. First, there is the lower-level programming kernels provided by CUDA in the C language. CUDA also provides two higher-level application programming interfaces (APIs), the first being Thrust, which provides strong support for lambda functions, and the second one is CUB, which provides a fast software component for processing data on the GPU. This work implements four variants of the filter using different programming strategies, one based on GPU-kernel \cite{navarro_hitschfeld-kahler_mateu_2014}, two based on Thrust, and one based on CUB. The code for these implementations is available at \href{https://github.com/rcarrascoc/GPU-2D-Convex\_Hull\_Filter}{https://github.com/rcarrascoc/GPU-2D-Convex\_Hull\_Filter}.

We define three functions for the filtering in algorithm \ref{algo}. The first function corresponds to finding the Eight-side polygon phase as \texttt{findingPolygon} where the input is a set of points, and the output is all extremes points and corners. The second function is \texttt{buildingFilter} which takes all extreme points and marks which points are candidates to the hull. The third function is the compaction of the input using the flag obtained from the filter in a non-sparse matrix, leaving in the output of this function only the hull candidate points. Finally, the convex hull is computed using any algorithm available on the-state-of-the-art with the output of the previous function. 

\begin{algorithm}
    \footnotesize
    \caption{Filter}\label{algo}
    \begin{algorithmic}[1]
    \Require Set of points $S$ 
    \Ensure Set of points candidates to the hull
        \State $\{left, top, right, down, c1, c2, c3, c4 \}$ $\leftarrow$ \textsc{findingPolygon($S$)} 
        \State $bit\_vector\_flag$ $\leftarrow$ \textsc{buildingFilter($S,left,top,right,c1,c2,c3,c4$)} 
        \State $filtered\_set$ $\leftarrow$ \textsc{compactingFilteredPoints($bit\_vector\_flag$)} 
        \State $output$ $\leftarrow$ \textsc{convexHull\_algorithm($filtered\_set$)} 
    \State \Return $output$
    \end{algorithmic}
\end{algorithm}

The naming convention for the variants means that the first word, textit{Thrust} or \textit{CUB}, indicates whether the method of finding the extreme points is using parallel reductions in a GPU kernel or through the Thrust or CUB functions, respectively. The second word (\textit{Scan}, \textit{Copy}, or \textit{Flagged}) refers to a specific feature that differentiates it from the others during compaction. This section is divided into three sub-sections that describe the method used to find the extreme points of the first word of the method, and each sub-section has its respective sub-subsection that describes the technique used for compaction according to the second word. These techniques are not interchangeable with each other, giving as the only combinations those described below.

\subsection{GPU Implementation \textit{(kernel)}}
The basic idea of this filter is to use only custom GPU kernels and techniques of the GPU programming, this implementation involves getting all extreme points (leftmost, rightmost, highest, and lowest) using parallel reduction techniques based on cuda-shuffle operations \cite{harris2007optimizing, harris_2005} to find the minimum and maximum point coordinates of each axis. Once the extreme points are obtained, the corners are calculated using parallel reductions again to find the minimum distances from Manhattan to each corner of the bounding box.

\subsubsection{Variant \#1: \texttt{proposed-filter}}
\label{subsec:stream}

To find the candidate points for the hull, we use a GPU kernel to check if the points are inside or outside the eight-sided polygon constructed from the extreme points and corners. The result of this check is a vector of bits of size $n$ (the same size as the number of input points), where bits marked with $1$ indicate that the point is a candidate for the hull, and bits marked $0$ indicate that the point has been discarded from the convex hull computation.

\begin{figure}[htb]
    \centering
    \includegraphics[scale=0.65]{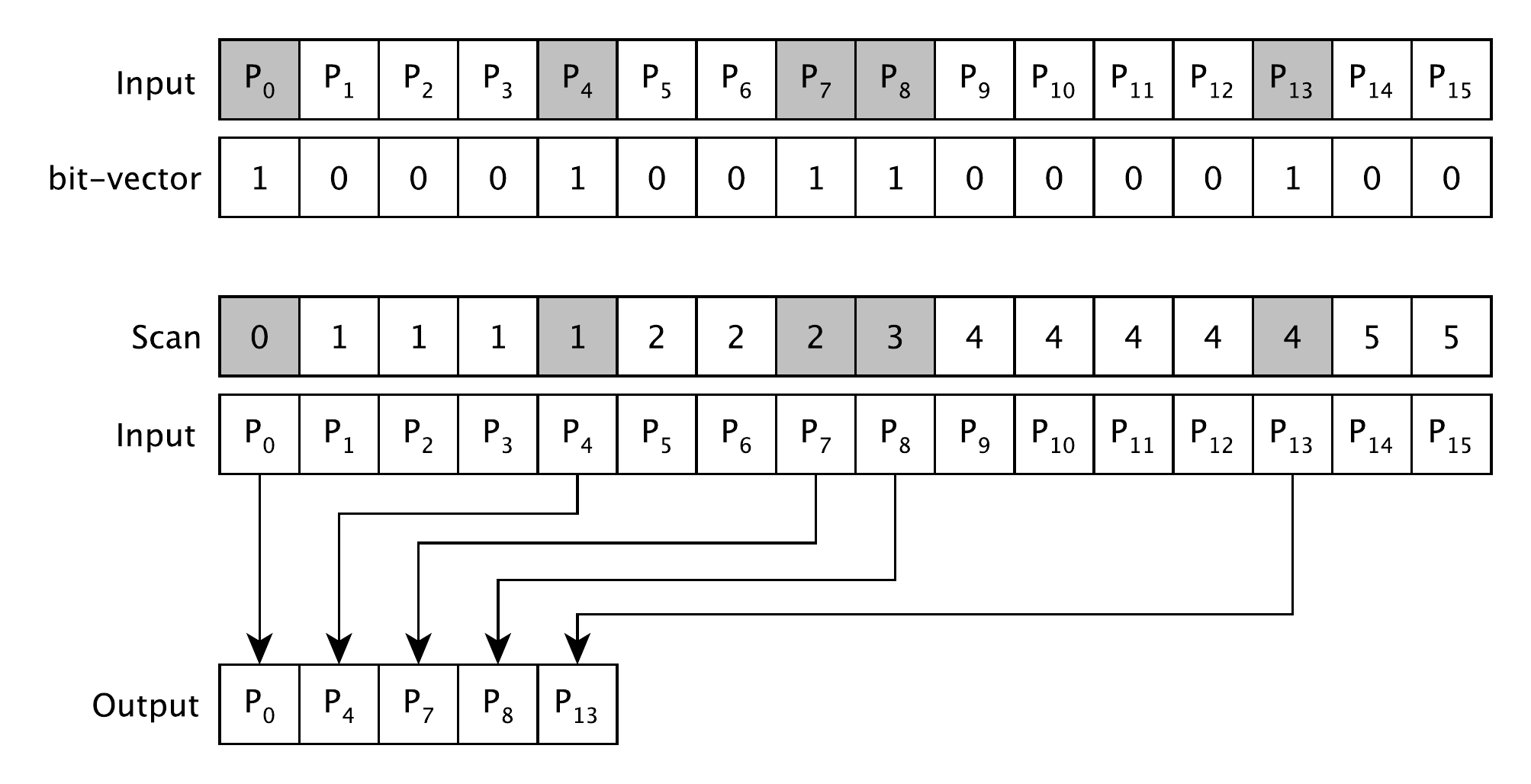}
    \caption{The illustration shows the underlying idea of parallel \textit{scan} compaction. The input points that we want to preserve are marked in gray, setting a $1$ at each gray cell. Then the scan of each combined $1$ is calculated.}
    \label{fig_scan_compaction}
\end{figure}

So now we have a sparse vector that identifies whether a point is a candidate for the hull or not. However, it is necessary to reduce the input for the convex hull to a reduced list of points. Since simultaneous global memory access in parallel computing is not a trivial problem, there are solutions that take logarithmic time. 
Figure \ref{fig_scan_compaction} shows a strategy that copies all items marked as $1$ and discards those marked as $0$. Following this idea, the location of each element marked with $1$ is given by the sum accumulation (\textit{scan}) \cite{42122} relative to the first point. This method requires three steps: filtering, scanning, and scattering, and it takes logarithmic time.

\begin{itemize}
\item The first pass generates a temporary vector in which elements that pass the filter are considered candidates for the convex hull and are set to $1$, while the other elements are set to $0$, as \textit{scan} input.
\item The second step performs a partial sum (\textit{scan}) on the bit vector generated in the previous step. Each element that passes the filter and is a candidate for the convex hull adds a position in the scan accumulation.
\item The third step places each point that is a candidate for the convex hull in the output array.
\end{itemize}

This work uses scanning techniques that take advantage of the tensor cores of modern GPUs, this technique can be divided into three main steps:
\begin{itemize}
    \item Segmented scanning: Tensor cores are used to perform a segmented scan \cite{ScanTC} on the temporary vector generated in the previous step. This step generates the new locations for each point with respect to each segment.
    \item Global scanning: using the most significant values of each segment is computed the scan of all segments using CUDA cores. This scan says the location of each segment with respect to the global position. 
    \item Compaction: Using the new locations generated in the previous step, the points that are candidates for the hull are compacted into the output array. Where the final location is given the $segment[pos]+global[pos/num\_segment]$.
\end{itemize}
This technique is highly efficient and can handle large arrays, allowing for faster compaction compared to other libraries such as Thrust and CUB.

\subsection{Thrust Implementations}
The Thrust API provides an easy-to-use high-level interface for parallel programming on CUDA-enabled GPUs. It offers a wide range of functionalities such as scan, sort, minimum-maximum reductions, data transformations, and array compaction, which have been widely used in previous works on convex hull algorithms in GPU \cite{cudach2d, cudach3d, mei2012}. However, it should be noted that the maximum size of vectors that can be processed with Thrust is $2^{30}$ points, which may limit its use in cases where a larger point set is required. In contrast, the kernel implementations are only limited by the available GPU memory, providing more scalability in terms of input size.

API Thrust offers the \texttt{min\_element} and \texttt{max\_element functions}, which return the indices of the minimum and maximum elements of a list, respectively. To find the extreme points, a max-min reduction must be performed on each axis coordinate. To find the corners, the Manhattan distances between each point and each corner must be computed, followed by a reduction. Finally, a function of transformation is used to check if each point belongs to the hull, using a bit vector. Two strategies for compacting the output can also be used, taking advantage of the reductions provided by Thrust.

\subsubsection{Variant \#2: \texttt{thrust-scan}}
This variant uses the same strategy as the scan variant (subsection \ref{subsec:stream}) but utilizes the \texttt{exclusive\_scan} function from Thrust to obtain the partial sum (\textit{scan}) from a bit vector generated during the filter stage. However, it is necessary to transform the vector resulting from the prefix sum into a compatible array stored in GPU memory, which is done by scattering the correct addresses of each point in a call to the GPU kernel. This variant takes advantage of the efficient reduction offered by Thrust. However, it incurs an additional computational cost when casting a Thrust vector to a GPU array, which can be avoided, as described in the next variant, by using a slower operation for compaction.

\subsubsection{Variant \#3: \texttt{thrust-copy}}
The \texttt{copy\_if} function provided by Thrust allows copying elements of an initial array to a new, smaller array based on a given condition. In this scenario, the condition is determined by the filter, where a point is retained if it is marked as $1$ by the filter and discarded if it is marked as $0$. While the \texttt{copy\_if} function performs well when the filter discards a significant number of points, it may become computationally expensive when the filter retains a large number of points.

\subsection{CUB Implementations}
CUB offers cutting-edge, reusable software components for various layers of the CUDA programming model. It includes three levels of CUB primitives:

\begin{itemize}
    \item Warp-wide "collective" primitives that operate within a single warp.
    \item Block-wide "collective" primitives that operate across all threads within a block.
    \item Device-wide primitives that operate across all threads and blocks on the device.
\end{itemize}
The first two levels correspond to operations that take place within a kernel, and the last level corresponds to operations on the host.

CUB provides device-wide primitive functions called \texttt{ArgMax} and \texttt{ArgMin} to find the maximum and minimum, respectively, represented as a pair of key-value, where the key corresponds to the index of the point, and the value corresponds to the coordinate value. 
CUB also provides a function for the \texttt{scan}, however, during experimentation it does not shows better performance than the other implemented variants.


\subsubsection{Variant \#4: \texttt{cub-flagged:}}
DeviceSelect offers device-wide, parallel operations for compacting selected items from sequences of data stored within device-accessible memory. These operations apply a selection criterion to selectively copy items from a specified input sequence to a compact output sequence. It uses in terms of the programming language, the \texttt{d\_flags} sequence to determine which items from \texttt{d\_in} to copy into \texttt{d\_out}, and the total number of selected items is written to \texttt{d\_num\_selected\_out}. This allows for efficient and parallel compaction of data on the device.

\section{Study cases}
\label{sec_study}
The number of filtered points is strongly dependent on the point distribution in the plane, where if the points are concentrated at the center of the points cloud, the number of filtered points is greater than if all the points are at the edge of the cloud.
This work explores three different interesting scenarios to study the behavior of the proposed approach, (1) a normal distribution, (2) a perfect circumference, and (3) a displaced circumference.

\begin{figure*}
     \centering
     \begin{subfigure}[b]{0.49\textwidth}
         \centering
         \includegraphics[width=\textwidth]{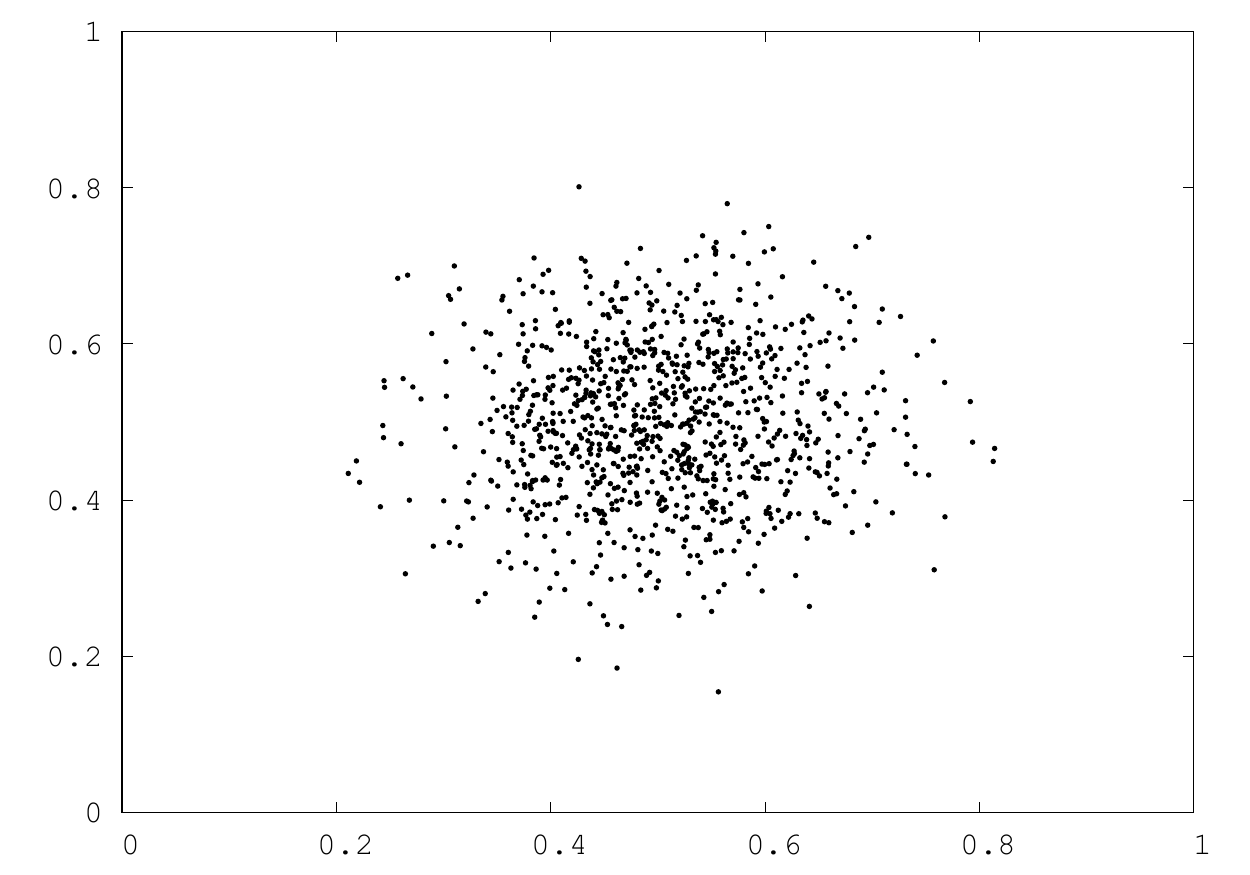}
         \caption{Normal distribution.}
    \label{fig:plot_norm}
     \end{subfigure}
     \hfill
     \begin{subfigure}[b]{0.49\textwidth}
         \centering
    \includegraphics[width=\textwidth]{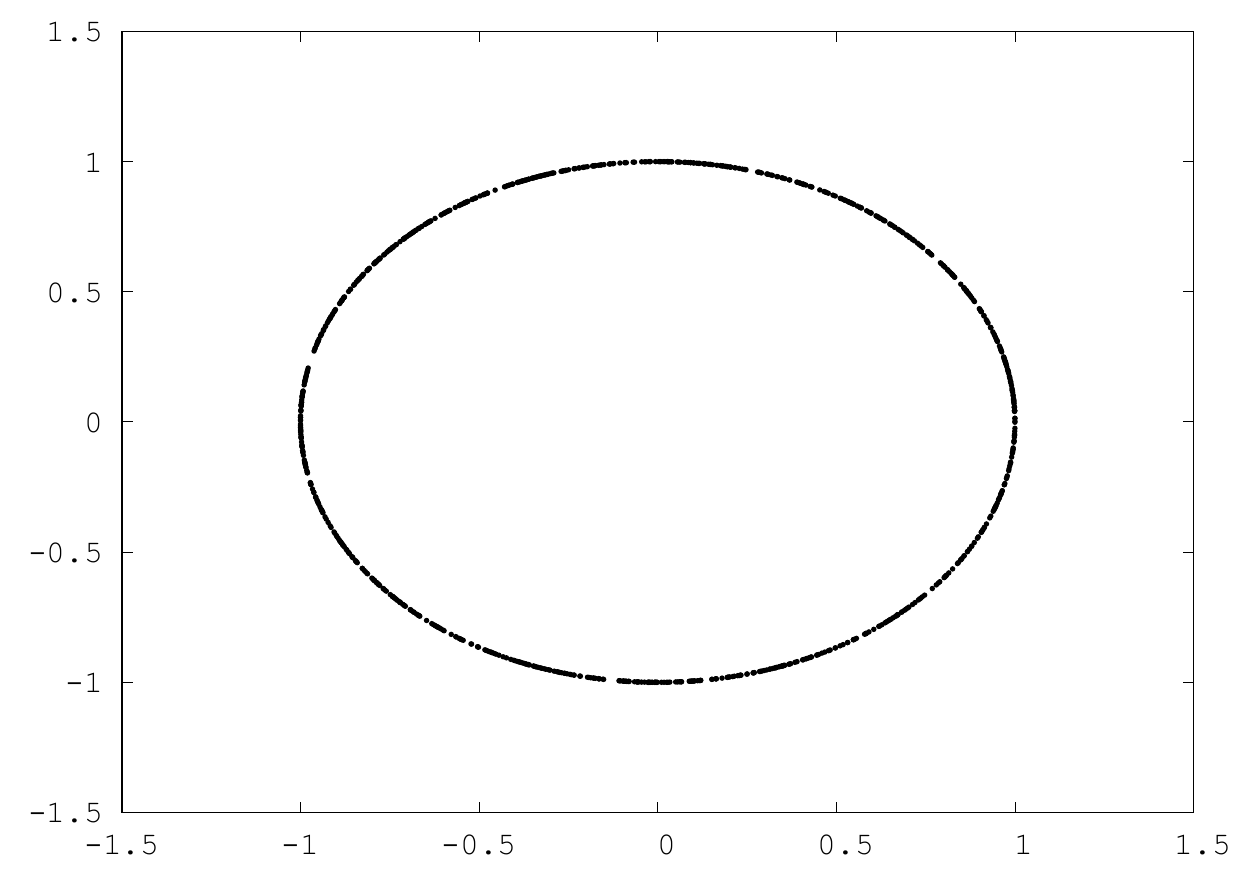}
         \caption{Circumference (worst case).}
    \label{fig:plot_circ}
     \end{subfigure}
        \caption{Graphic representation of the distributions, the figure on the left side corresponds to a normal distribution with a mean $1$, and a standard deviation of $0.1$; next to it is a circumference centered on the origin of radius $1$}
        \label{fig:plot_plot}
\end{figure*}

\begin{figure*}
     \centering
     \begin{subfigure}[b]{0.45\textwidth}
         \centering
         \includegraphics[width=\textwidth]{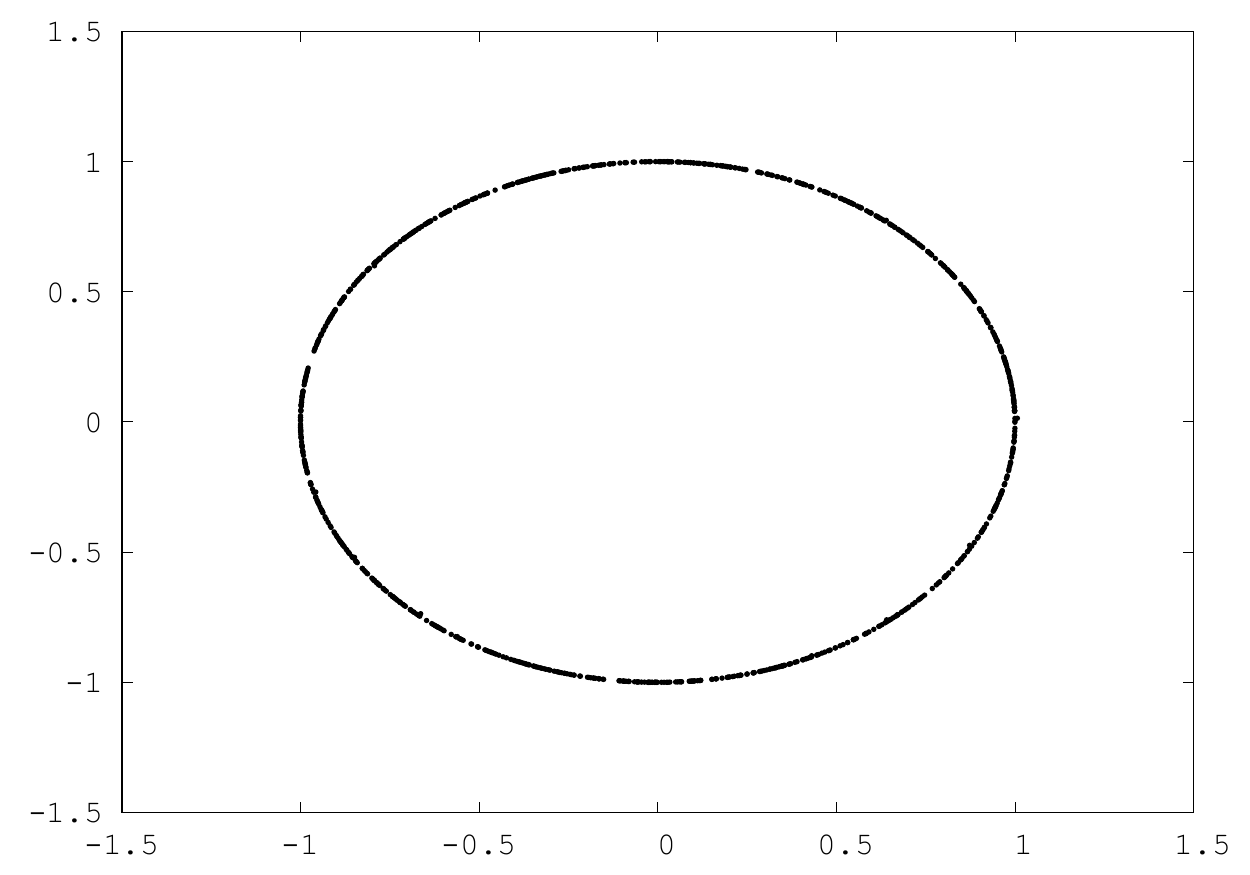}
         \caption{$p=0.01$}
         \label{fig:plot001}
     \end{subfigure}
     \hfill
     \begin{subfigure}[b]{0.45\textwidth}
         \centering
         \includegraphics[width=\textwidth]{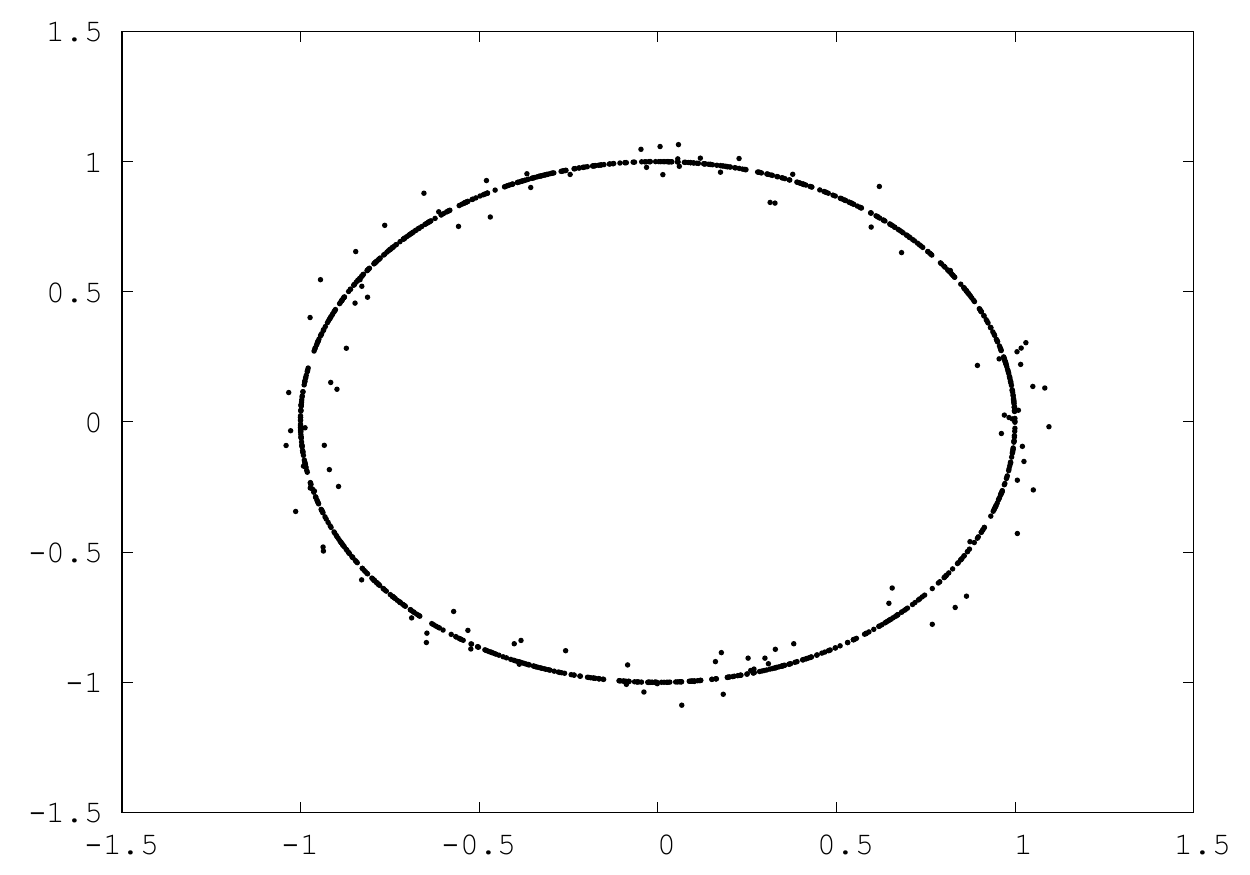}
         \caption{$p=0.10$}
         \label{fig:plot010}
     \end{subfigure}
     \vfill
     \begin{subfigure}[b]{0.45\textwidth}
         \centering
         \includegraphics[width=\textwidth]{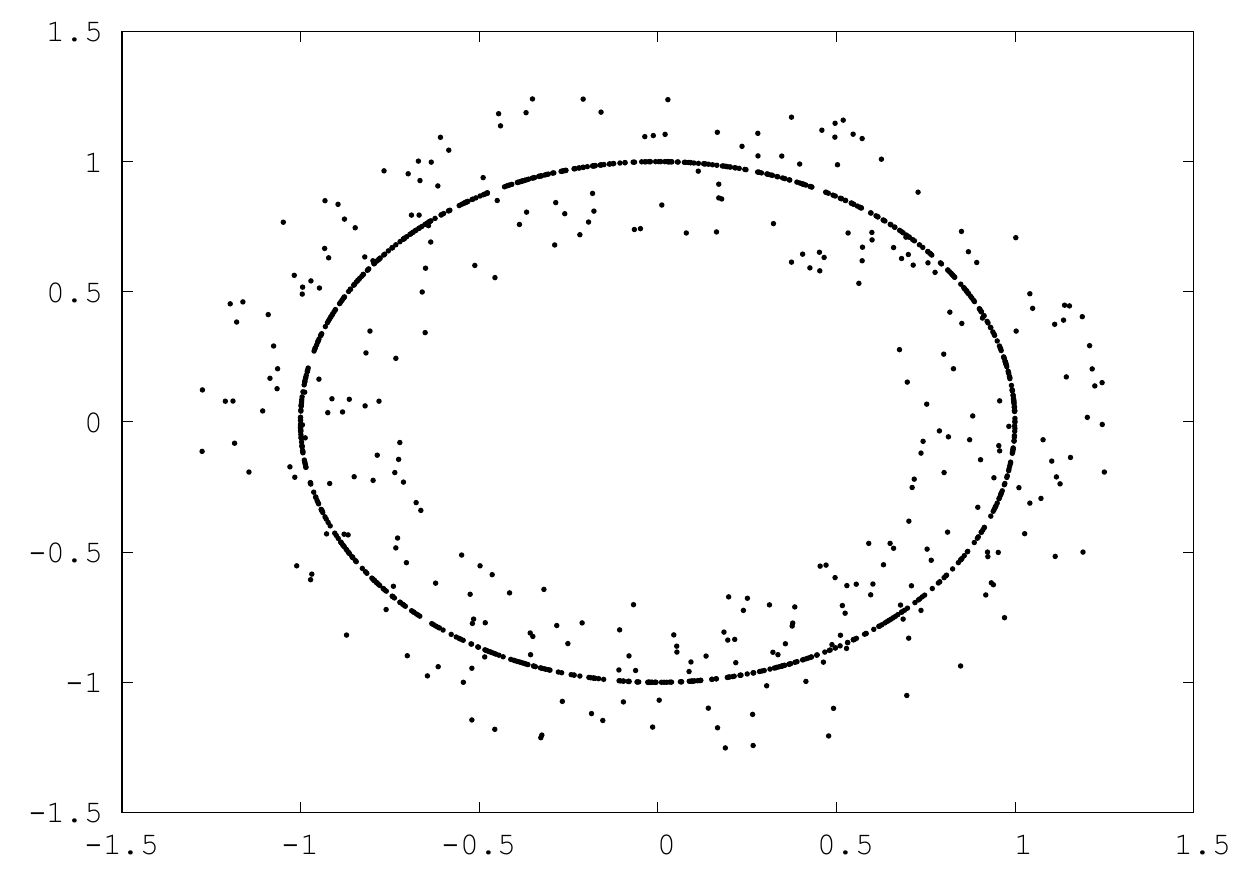}
         \caption{$p=0.30$}
         \label{fig:plot010}
     \end{subfigure}
     \hfill
     \begin{subfigure}[b]{0.45\textwidth}
         \centering
         \includegraphics[width=\textwidth]{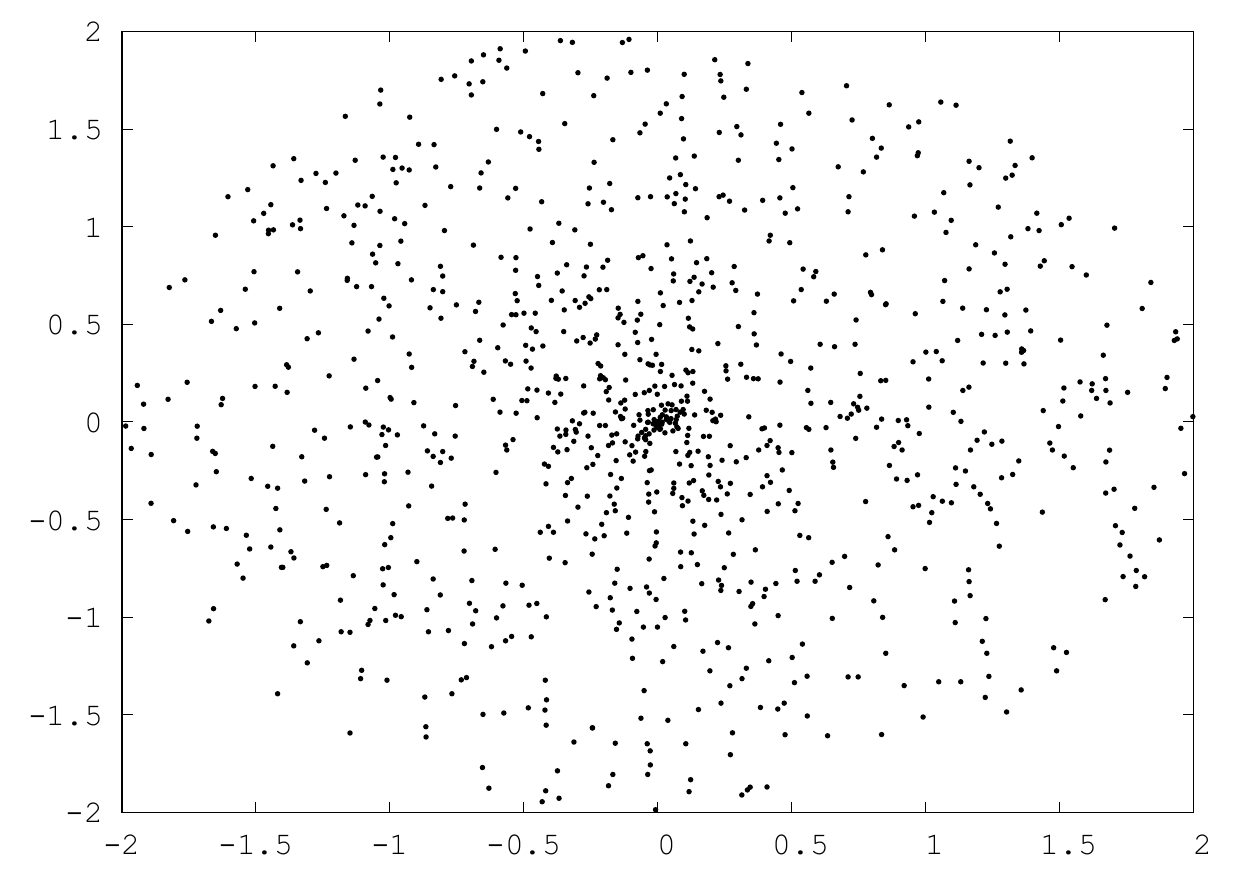}
         \caption{$p=1.00$}
         \label{fig:plot100}
     \end{subfigure}
        \caption{The intermediate-case test with randomly selected displacement at different values.}
        \label{fig:plot_inter}
\end{figure*}

\begin{itemize}
    \item [1.-  ] \textbf{Normal Distribution - Normal case:}
    A normal distribution is usually found in many problems such as physical phenomena, human behavior, sciences, and other disciplines, as these events naturally tend to behave similarly to a Gaussian distribution, as in Figure \ref{fig:plot_norm}. This test generates $n$ random points normally distributed in the plane with $\mu = 0.5$ and $\sigma^2 = 0.1$. Some convex hull algorithms output-sensitive like \textit{QuickHull} take advantage of this type of case.
    
    \item [2.-  ] \textbf{Perfect Circumference - Worst case:}
    Figure \ref{fig:plot_circ} shows the case when all points are part of a circumference, and so all points are in the convex hull.  Since no point is filtered, this is the worst case for the proposed algorithm. Unlike the normal distribution, some algorithms are output-sensitive, like the \textit{Gift Wrapping} algorithm, which is ineffective. In this case, it is recommendable to use an algorithm to guarantee a good performance in $O(n \log n)$.
    
    \item [3.-  ] \textbf{Displaced Circumference - Intermediates cases:}
    This test permits the model of point distributions with  more or fewer points  belonging to the hull. Here we look for the tipping point where the algorithms start to perform well. The test is created by generating n points on a circumference (or very close to it) centered at the origin with radius $r = 0.25$. The test offers a $p$ parameter to be chosen, which produces a displacement probability for each point in the range $[r-rp,r+rp]$. This displacement may move the point inwards to the center or outwards the circumference, making a band of points surrounding the circumference. Figure \ref{fig:plot_inter} shows the displacement for different $p$ values.
\end{itemize}

\section{Experiments and Results}
\label{sec_experiment}
We implemented the filtering algorithm in C++ with -O3 optimization for the CPU part, and CUDA with NVCC 11.4.2 for the GPU part. Our implementation uses single-precision floating-point arithmetic (FP32). We compare its performance with the CGAL 5.5 library, one of the most widely used and cited libraries. We perform all experiments on the Patag\'on supercomputer \cite{patagon}, which has one  Nvidia DGX A100 GPU node, 2x AMD EPYC 7742 CPU (2.6GHz, 64-cores, 256MB L3 cache), 1 TB RAM DDR4-3200Hz, and 8x Nvidia A100 GPUs 40GB. The experiments only use one A100 GPU. For each variant, the following metrics were collected: filtering time, percentage of points filtered and runtime of the variants developed in this work. 

\begin{figure*}
     \centering
     \begin{subfigure}[b]{0.45\textwidth}
         \centering
         \includegraphics[width=\textwidth]{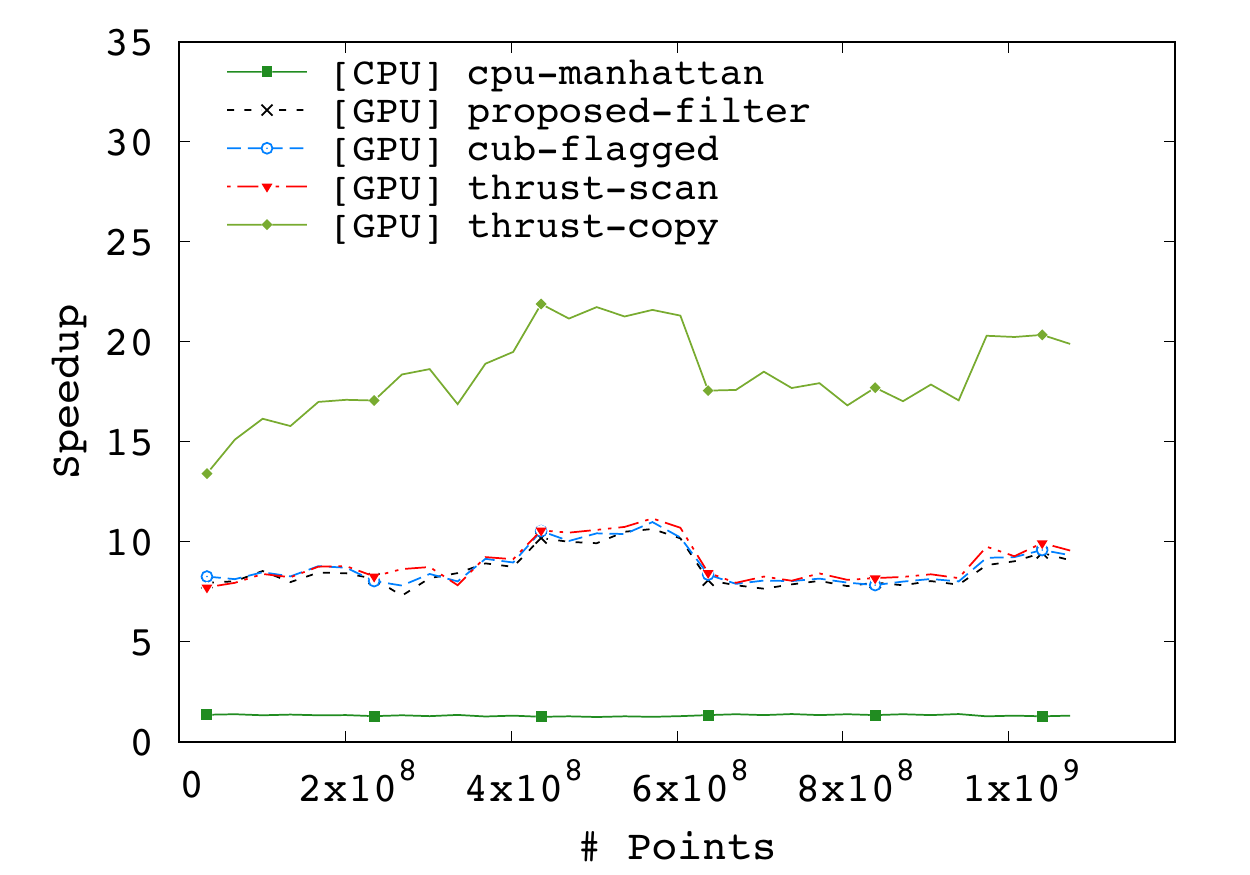}
         \caption{Filter for a normal distribution}
         \label{fig:norm_filter}
     \end{subfigure}
     \hfill
     \begin{subfigure}[b]{0.45\textwidth}
         \centering
    \includegraphics[width=\textwidth]{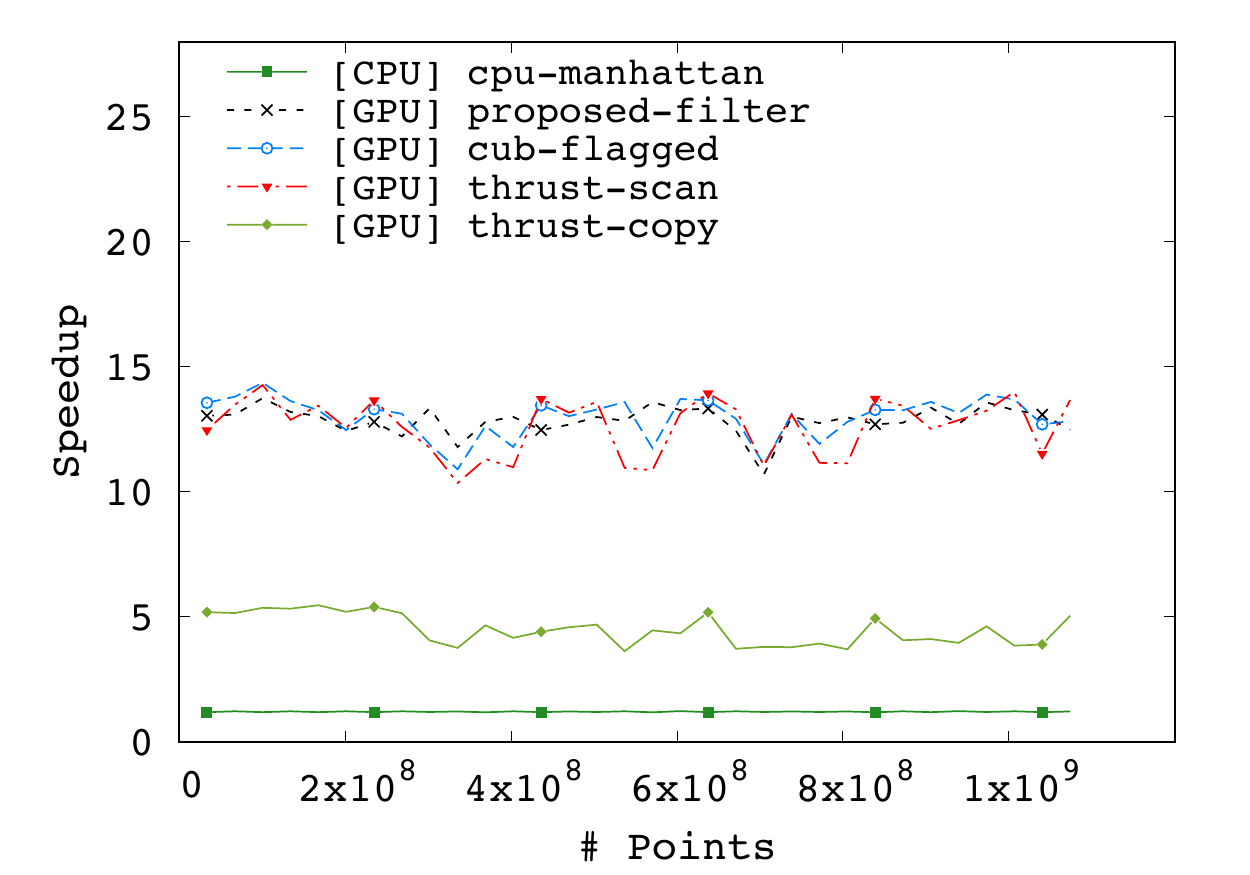}
         \caption{Filter for a circumference}
         \label{fig:circ_filter}
     \end{subfigure}
        \caption{Speedup over \texttt{cpu-euclidean} filter of only the preprocessing filtering phase for a normal and circumference distribution, between $2^{25}$ and $2^{30}$ \# points.}
        \label{fig:filter_speedup}
\end{figure*}

We benchmarked the performance of the filter variants by comparing them to their CPU counterparts (using Euclidean and Manhattan distances to find extreme points) in figure \ref{fig:filter_speedup}. As shown in Figure \ref{fig:norm_filter}, the fastest implementation for a normal distribution is \texttt{thrust-copy}, achieving a speedup of $23 \times$ over \texttt{cpu-euclidean} filter. However, for a circumference distribution, it is the slowest. The other implementations have similar performance, in figure \ref{fig:circ_filter}, achieving a speedup of $11 \times$ in normal distribution and $14 \times$ in circumference distribution, and are faster than both \texttt{thrust-copy} and CPU variants. It's important to note that all GPU variants are sensitive to the number of filtered elements.

\begin{figure*}
     \centering
     \begin{subfigure}[b]{0.45\textwidth}
         \centering
         \includegraphics[width=\textwidth]{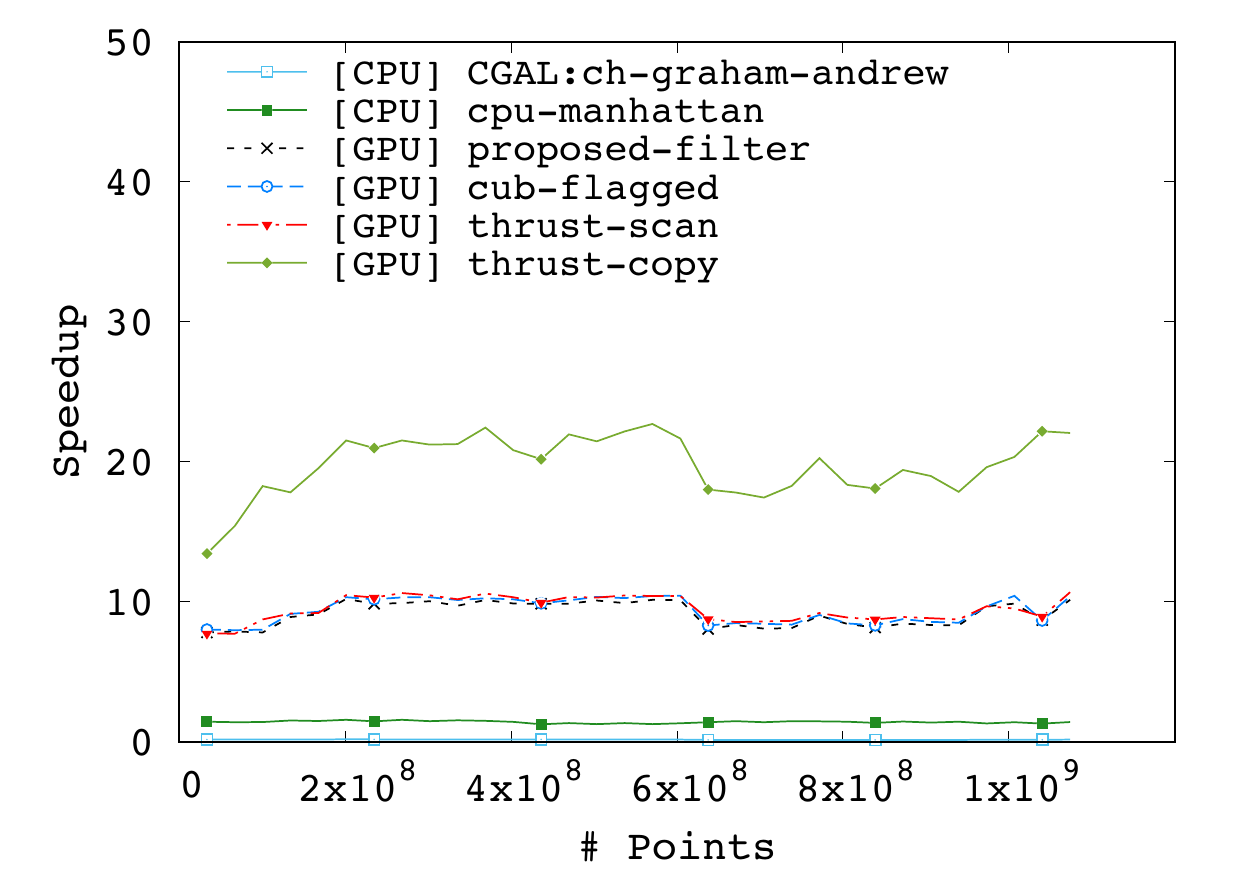}
         \caption{Normal distribution.}
    \label{fig:ch_norm}
     \end{subfigure}
     \hfill
     \begin{subfigure}[b]{0.45\textwidth}
         \centering
    \includegraphics[width=\textwidth]{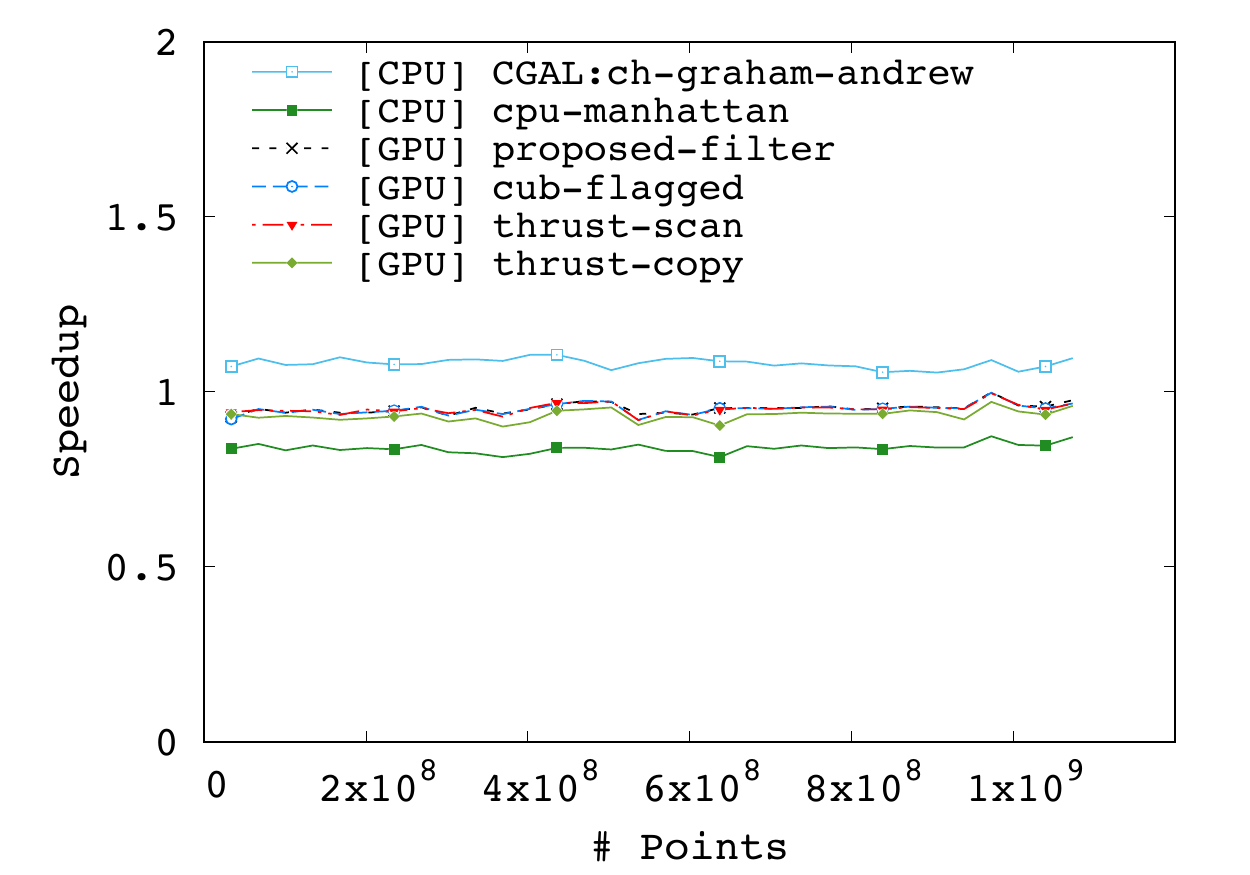}
         \caption{Circumference.}
    \label{fig:ch_circ}
     \end{subfigure}
        \caption{Speedup of filter + \texttt{CGAL:convex-hull-2} over a \texttt{CGAL-convex-hull-2} without filter for a normal and circumference distribution, respectively, between $2^{25}$ and $2^{30}$ \# points.}
        \label{fig:ch_speedup}
\end{figure*}

Figure \ref{fig:ch_speedup} shows the acceleration of all the variants described in this work using the fastest implementation provided by CGAL (\texttt{CGAL:convex-hull-2}) to compute the hull, the speedup is calculated with respect to \texttt{CGAL:convex-hull-2} without a filter. Note that \texttt{CGAL:convex-hull-2} corresponds to a hybrid algorithm that chooses between an $O(hn)$ or $O(n \log n)$ algorithm technique depending on the distribution calculus.
The acceleration time for a normal distribution shown in Figure \ref{fig:ch_norm} indicates that \texttt{thrust-copy} variant offers a significant acceleration of $23 \times$. At the same time, the other variants achieve a speedup of $10 \times$. However, for a circumference distribution, the acceleration with any preprocessing algorithm is not possible given the nature of the problem, as shown in Figure \ref{fig:ch_circ}. Unlike the CPU variants, the GPU variants only have a slight overhead. 

\begin{figure*}
     \centering
     \begin{subfigure}[b]{0.45\textwidth}
         \centering
         \includegraphics[width=\textwidth]{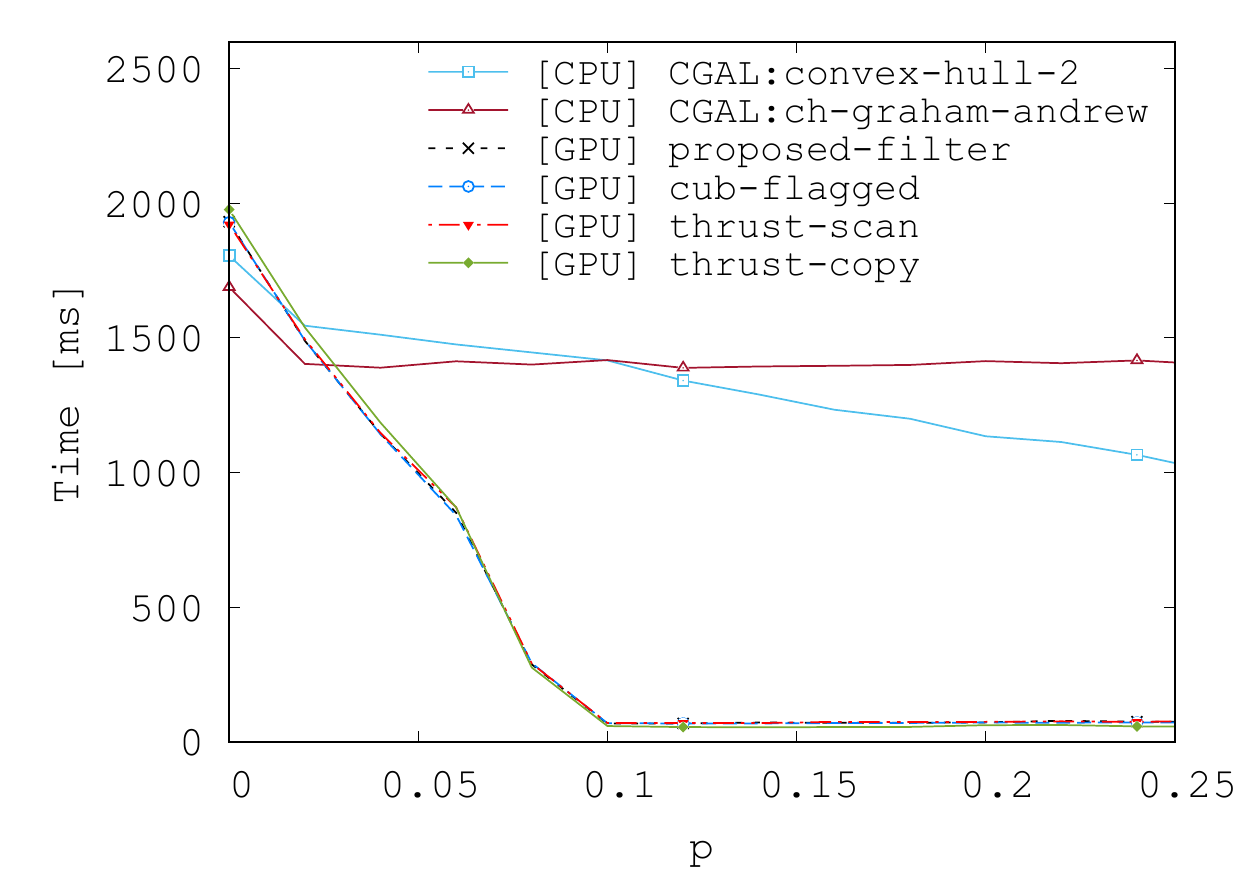}
         \caption{Size of $10^7$ points}
    \label{fig:ch7}
     \end{subfigure}
     \hfill
     \begin{subfigure}[b]{0.45\textwidth}
         \centering
    \includegraphics[width=\textwidth]{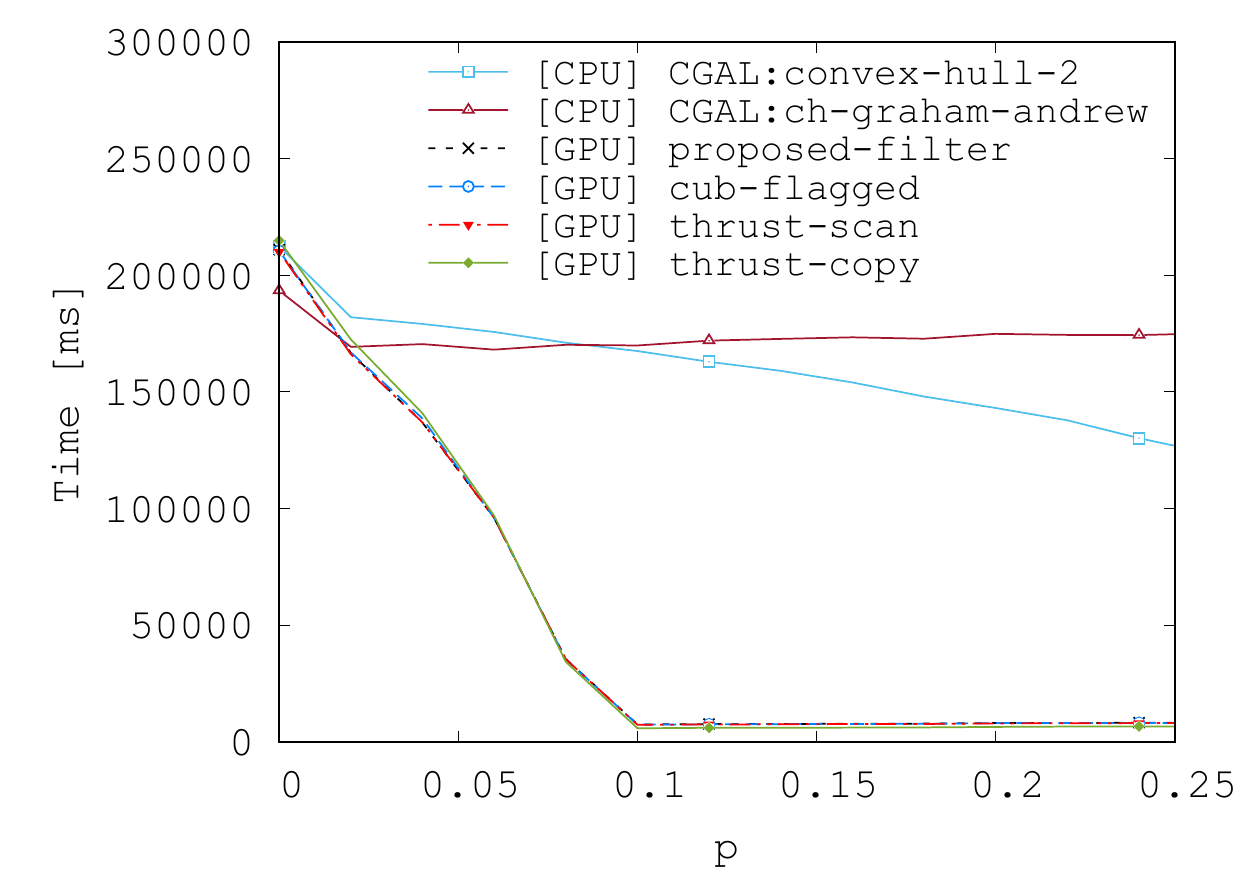}
         \caption{Size of $10^9$ points}
    \label{fig:ch9}
     \end{subfigure}
        \caption{Running time of the convex hull algorithm for a displaced circumference (intermediate case) varying $p$ between $[0, 0.25]$, and fixed size.}
        \label{fig:ch0}
\end{figure*}

\begin{figure*}
     \centering
     \begin{subfigure}[b]{0.45\textwidth}
         \centering
         \includegraphics[width=\textwidth]{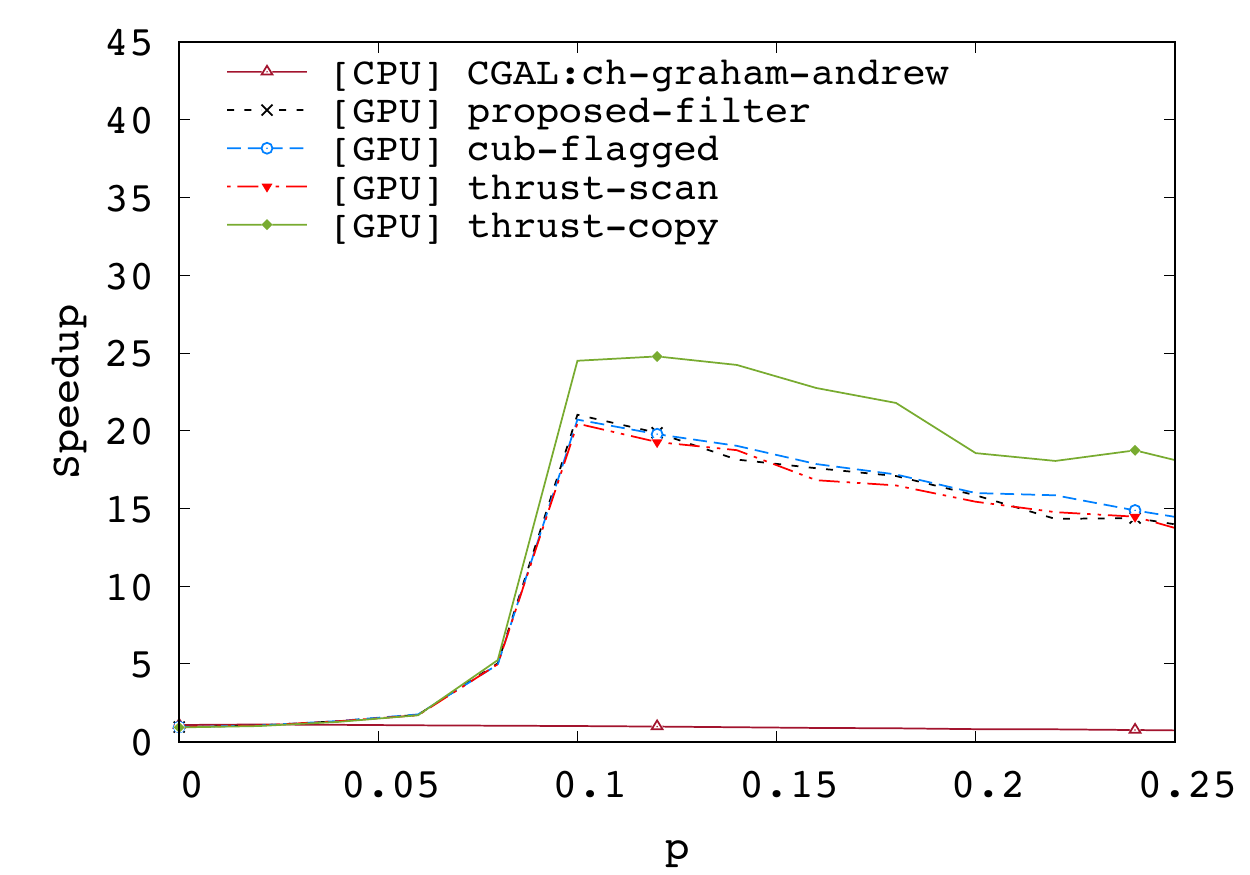}
         \caption{Size of $10^7$ points}
     \end{subfigure}
     \hfill
     \begin{subfigure}[b]{0.45\textwidth}
         \centering
    \includegraphics[width=\textwidth]{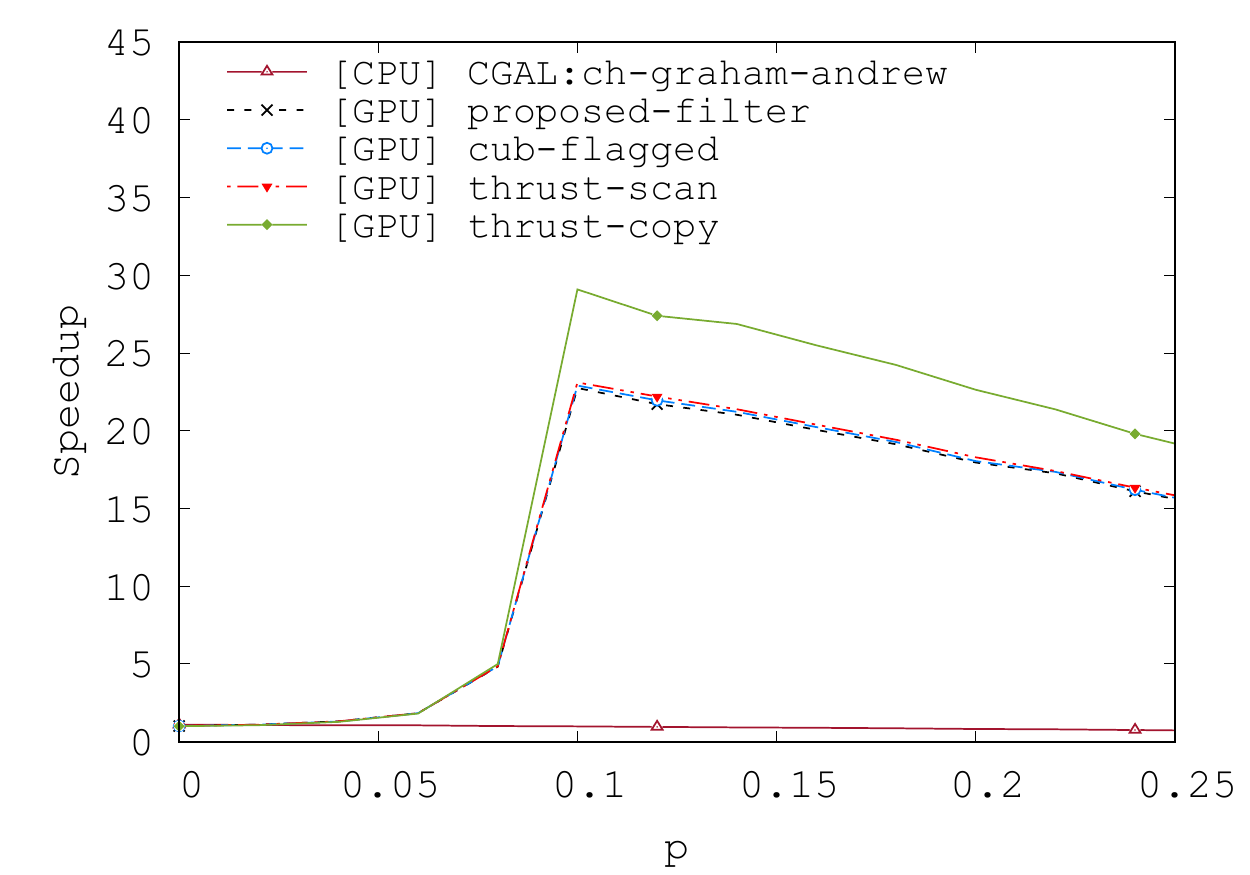}
         \caption{Size of $10^9$ points}
     \end{subfigure}
        \caption{Speedup of the convex hull algorithm over \texttt{CGAL:convex-hull-2} for a displaced circumference (intermediate case) varying $p$ between $[0, 0.25]$, and fixed size.}
        \label{fig:ch0_speedup}
\end{figure*}

Figure \ref{fig:ch0} illustrates that a minimum number of points is necessary for the filtering to speed up the computation of the lock. Figure \ref{fig:ch7} shows the time it takes to filter a set of small points. Instead, figure \ref{fig:ch9} shows the filtering time for a large set of points. As can be seen, the behavior of both data sets is different, where the table \ref{tab:filterig} indicates that for a small data set, it is required to filter $p < 0.03$ so that the filtering speeds up the computation of the closure, and for large data sets $p < 0.01$ is required, which results in the removal of about $6\%$ of points. Furthermore, it is crucial to consider that the highest acceleration is achieved with $p=0.1$, as observed in both plots of figure \ref{fig:ch0}, where a valley is reached with the maximum acceleration. Finally, we can see the speedup for both distributions in Figure \ref{fig:ch0_speedup}, where the fastest variant (\texttt{thrust-copy}) reaches up to $30 \times$ over \texttt{CGAL:convex-hull-2} for a large data set in $p=0.1$.

\begin{table}[ht!]
\begin{tabular}{c|c|c|c|c|} \cline{2-5}
& \multicolumn{2}{c}{$n=10^7$ points} & \multicolumn{2}{|c|}{$n=10^9$ points} \\ \hline
\multicolumn{1}{|c|}{$p$} & filtered \% & Speedup & filtered \% & Speedup \\  \hline
\multicolumn{1}{|c|}{0.00} & \textcolor{red}{0.01} &  \textcolor{red}{0.88} & \textcolor{red}{0.01} &  \textcolor{red}{0.92} \\ 
\multicolumn{1}{|c|}{0.02} & \textcolor{red}{13.08} &  \textcolor{red}{0.94} & \textcolor{blue}{13.04} &  \textcolor{blue}{1.02}\\ 
\multicolumn{1}{|c|}{0.04} & \textcolor{blue}{28.58} &  \textcolor{blue}{1.21} & 28.69 & 1.24\\
\multicolumn{1}{|c|}{0.06} & \textcolor{black}{48.48} & 1.62 & 48.51 & 1.75\\ 
\multicolumn{1}{|c|}{0.08} & 81.88 & 4.83 & 81.71 & 4.82 \\ 
\multicolumn{1}{|c|}{0.10} & 97.16 & 20.46 & 97.17 & 23.42\\\hline
\end{tabular}
\caption{Average percentage of discarded points at the filtering stage in the algorithm for a displaced circumference, and speed up the fastest variant with respect to \texttt{CGAL:ch-graham-andrew}. }
\label{tab:filterig}
\end{table}

\section{Discussion and Conclusion}
\label{sec_conclu}
A complete performance evaluation of the most recent GPU filtering techniques has been presented in this work to solve the 2D Convex hull problem. Although there are GPU implementations that use custom CUDA codes, and some others that use the Thrust library, this  work contributes in comparing three main ways of doing GPU filtering; custom CUDA kernel, Thrust and CUB libraries, while highlighting the advantages and disadvantages of each.

The proposed filtering method builds an eight-vertex polygon, where all interior points are discarded, and utilizes parallel min-max reduction and the Manhattan distance to compute the corners. Manhattan distance simplifies the computation and eliminates the need for square root operations. The experimental results using randomly generated points with a normal distribution in 2D space indicate that the proposed method accelerates the computation of the convex hull by $30 \times$ with respect to \texttt{CGAL:convex-hull-2} function provided by CGAL.

In cases where all points lie on the convex hull, such as in circular distributions, the proposed approach does not offer any benefit as the filtering algorithms do not remove points, a characteristic of all state-of-the-art algorithms. However, the cost of running the filtering algorithm on the GPU is minimal (less than $6\%$) compared to running it on the CPU. This means that using the GPU-based filtering process only results in a small increase in computation time for the convex hull. In some cases, depending on the application, it may be acceptable to bear the cost in exchange for the potential speedup in other scenarios.

An positive aspect of this work is the scalability of the algorithm. Thrust and CUB provide a fast and user-friendly API for developing parallel algorithms. However, its functions for computing the minimum and maximum points are unsuitable for processing large data sets (more than $2^{30}$ points). CGAL also has a maximum limit of $2^{30}$ points. On the other hand, the proposed GPU-kernel implementation is easily scalable and can process a larger number of points as long as the graphic memory allows it. Moreover, the GPU-kernel implementation achieves the same performance as the variants based on libraries.

The experimental results indicate that this methodology is efficient in most cases, and even in the worst-case scenario, the filtering effort has a small performance penalty compared to a traditional Convex Hull computation. Moreover, the worst-case scenario is in most cases very unlikely to occur.
As future work, it would be interesting to study a complete parallel convex hull algorithm on the GPU, utilizing the proposed preprocessing aproach and avoiding unnecessary data copying between the device and host memory. Additionally, accelerating the 3D convex hull and leveraging tensor and ray tracing cores for this task have become relevant topics.

\section*{Acknowledgment}
This research was supported by the Patag\'on supercomputer of Universidad Austral de Chile (FONDEQUIP EQM180042), and FONDECYT grants $\#1211484$, $\#1221357$.

\printcredits

\bibliographystyle{unsrt}

\bibliography{cas-refs}

\end{document}